\documentclass[aps,pra,reprint,twocolumn,superscriptaddress,nofootinbib,preprintnumbers]{revtex4-1}
\pdfoutput=1 

\usepackage{graphicx}
\usepackage{bm}
\usepackage{amsmath}
\usepackage{amssymb}
\usepackage{amsfonts}
\usepackage{amsthm}
\usepackage{mathtools}
\usepackage{hyperref}
\usepackage{braket}
\usepackage[normalem]{ulem}
\usepackage{wrapfig}
\usepackage{tikz}
\usepackage{dsfont}
\usepackage{comment}
\usepackage{lipsum}

\hypersetup{colorlinks=true,urlcolor=[rgb]{0,0,0.5},citecolor=[rgb]{0.5,0,0},linkcolor=[rgb]{0,0,0.4}}

\def\beq{\begin{equation}}
\def\eeq{\end{equation}}

\begin{document}
\title{Fractional Hall physics from large $N$ interacting fermions}

\preprint{UTWI-41-2023}

\author{Kristan Jensen}
\email{kristanj@uvic.ca}
\affiliation{\it Department of Physics and Astronomy, University of Victoria, Victoria, BC V8W 3P6, Canada}
\author{Amir Raz}
\email{araz@utexas.edu}
\affiliation{\it University of Texas, Austin, Physics Department, Austin TX 78712, USA}

\begin{abstract}
We solve models of $N$ species of fermions in the lowest Landau level with $U(N)$-invariant interactions in the $N\gg 1$ limit. We find saddles of the second quantized path integral at fixed chemical potential corresponding to fractional Hall states with filling $ \frac{p}{q}$ where the integers $p$ and $q$ depend on the chemical potential and interactions. On a long torus there are $q$ such states related by translation symmetry, and $SU(N)$-invariant excitations of fractional charge. Remarkably, these saddles and their filling persist as extrema of the second-quantized action at $N=1$. Our construction gives a first-principles derivation of fractional Hall states from strongly interacting fermions. 
\end{abstract}

\maketitle

\textit{Introduction.}~The fractional quantum Hall effect, first observed in~\cite{firstFQH}, is a relatively simple arena for strong coupling physics. Non-interacting electrons in an applied field develop a series of gapped states, completely filled Landau levels, with integer filling fraction. That picture is completely spoilt by interactions, which split the degeneracy of the Landau levels and leave behind a network of fractionally filled gapped states. There is good reason to believe that the states with filling fraction $0<\nu<1$ arise from interacting electrons in the lowest Landau level, for which the effective Hamiltonian is entirely composed of interactions. Beginning with Laughlin~\cite{Laughlin_1983}, a mature and successful theory of the fractional Hall effect has been developed~\cite{PhysRevLett.51.605,PhysRevLett.52.1583,PhysRevLett.63.199,Moore:1991ks,PhysRevB.44.5246}; see~\cite{RevModPhys.71.S298,RevModPhys.89.025005} for reviews. However, this existing theory is a bit roundabout, for example constructing trial wavefunctions for the many-body states ground states or positing an effective Chern-Simons-matter field theory, rather than directly diagonalizing the interacting Hamiltonian.

In contrast, the goal of this Letter is to develop a first-principles theory of the fractional Hall effect. We work with a second quantized, many-body description of interacting fermions in the lowest Landau level, which we then solve at finite interaction strength. To do so we introduce $N$ species of fermions, mandate that the interactions are $U(N)$ invariant, and take $N\gg 1$, so that we can expand in the small parameter $1/N$. This sort of approximation has a long history in theoretical particle physics beginning with~\cite{tHooft:1973alw}, where it has been used to uncover non-perturbative phenomena like chiral symmetry breaking, dynamically generated scales, confinement, and more. Usually a large $N$ approximation is expected to describe qualitative features of a finite $N$ system; in this setting, due to the quantized nature of fractional Hall states, we will see that certain large $N$ computations are applicable even at $N=1$. 

Besides solvability at finite interaction strength, there are two more advantages of a second quantized  approach worth noting. The first is that, rather than constructing trial wavefunctions for individual fractional Hall states, we are able to uncover a hierarchy of plateaux along with the transitions between them. The second, related to the first, is that the many-body problem is naturally phrased in the grand canonical ensemble. Working in the canonical ensemble of fixed particle number raises the question of how the filling fraction stays constant as the magnetic field is adjusted. While this can be explained with disorder, see for example \cite{tong2016lectures}, it is much simpler to work in the grand canonical ensemble of fixed chemical potential, wherein a plateau arises when the chemical potential lies in a gap of the spectrum and no disorder is required.

In this Letter we study spin-polarized fermions confined to the surface of a cylinder of radius $r$, subject to a perpendicular magnetic field pointing out of the cylinder. The fermions are projected to the lowest Landau level, and the filling fraction is determined by varying a chemical potential $\mu$. As we will see, our results are closely related to those of \cite{Tau_Thouless}, and and especially the thin torus limit of \cite{Bergholtz_QHTT}, though both these references primarily worked in the canonical ensemble with $r\to 0$.

The main takeaway of this Letter is that we can study the emergence of fractional Hall physics and hierarchies in a controlled way starting from a UV theory of interacting fermions. We believe this viewpoint complements existing methods, and contributes a more comprehensive understanding of the fractional Hall effect.

The rest of this Letter is organized as follows. We begin by writing down the models in question and solving them at leading order in large $N$. We find saddles of the second quantized description corresponding to abelian fractional Hall states at filling fractions $\frac{p}{q}$ which dominate at zero temperature with transitions between them. These solutions spontaneously break the magnetic translations along the cylinder by a discrete subgroup, similar to the TT solutions of \cite{Tau_Thouless,Bergholtz_QHTT}, and the Laughlin states on the cylinder \cite{Jansen_2008}. We then find the basic fractionally charged quasiparticle excitations, and discuss what aspects of our solution remain at $N=1$. We also discuss a way to get ``$N=1$'' physics, meaning states with filling fraction $\frac{p}{q}$ and anyons of charge $\frac{1}{q}$ (for $p$ and $q$ coprime), by projecting the large $N$ theory to its $SU(N)$ singlet sector.

\textit{Interacting fermions in the Lowest Landau level.}~We work in a second-quantized description and consider $N$ species of fermionic fields  $\psi^{\alpha}$ living on cylinder of radius $r$, subject to an external magnetic field $B$. Throughout this note we will work in the Landau gauge 
\begin{equation}
    \mathbf{A} = B x \hat{y}\,,
\end{equation}
where the $y$ direction is periodic: $y \equiv y + 2\pi r$ and $x\in\mathbb{R}$. This gauge choice preserves continuous translations in $y$ as well as a discrete magnetic translation $x\to x + \frac{1}{Br}$. As our gauge choice is already periodic in $y$, we can simply impose periodic boundary conditions on the fermions: $\psi^{\alpha}(x,y+ 2\pi r) = \psi^{\alpha}(x,y)$.\footnote{We can also study ``twisted'' boundary conditions, but these do not substantially affect our results.}

To project these fermions to the lowest Landau level we impose the constraint \cite{muruyama2006lectures}
\begin{equation}
    \bar{D} \psi^{\alpha} = \frac{1}{2}\left(\partial_x + Bx -i\partial_y\right) \psi^{\alpha} = 0\,,
\end{equation}
through $N$ Lagrange multiplier fields $\chi_{\alpha}$ similar to \cite{nguyen2014lowest}. The field theory describing interacting fermions in the lowest Landau level then has the action
\beq
	S= \int dt~ d^2 x\bigg( i \bar{\psi}_{\alpha} D_t \psi^{\alpha} + \bar{\chi}_\alpha\bar{D}\psi^\alpha + D \bar{\psi}_\alpha \chi^\alpha \bigg) + S_{\text{int}}\,,
\eeq
with $D_t\psi = (\partial_t - i \mu)\psi$. We take $S_{\text{int}}$ to be a translationally-invariant two-body interaction labeled by a potential $V$,
\begin{equation}
    S_{\text{int}} = -\frac{1}{N}\int dt d^2xd^2x' ~ \rho(t,\vec{x}) V(|\vec{x}-\vec{x}'|) \rho(t,\vec{x}') \,,
\end{equation}
with $\rho(t,\vec{x}) = \bar{\psi}_{\alpha}(t,\vec{x})\psi^{\alpha}(t,\vec{x})$. 

On the cylinder the linearly-independent solutions to the lowest Landau level constraint $f_n$ are labeled by an integer $n$ with
\begin{equation}
     f_n(x,y) = \left( \frac{B}{4\pi^3 r^2}\right)^{\frac{1}{4}}
      e^{\frac{i n y}{r} - \frac{B}{2} \left(x +\frac{n}{Br}\right)^2}\,.
\end{equation}
Thus upon integrating out the $\chi_{\alpha}$ we have
\begin{equation}
    \psi^\alpha(t,x,y) = \sum_{n\in\mathbb{Z}} c^\alpha_n(t)f_n(x,y)\,,
\end{equation}
and the path integral over $(\chi_{\alpha},\psi^{\beta})$ reduces to one over the $c^{\alpha}_n$ on which magnetic translations act as $n\to n+1$. The action for the $c_n^{\alpha}$ is that of a 1d fermionic chain
\begin{equation} \label{eq:action}
\begin{split}
    S= \int dt\left( i \bar{c}^n_\alpha D_t c^\alpha_n - V_{n k }^{m l}~ \bar{c}_\alpha^{n} c_{m}^\alpha \bar{c}_\beta^{k} c_{l}^\beta\right)\,,
\end{split}
\end{equation}
where the interaction matrix $V_{n k }^{m l}$ is given by
\begin{equation}
    V_{n k }^{m l} = \int d^2xd^2x' ~ \bar{f}_n(x) f_m(x) V(|x-x'|)  \bar{f}_k(x') f_l(x')\,.
\end{equation}
We have in mind attractive interactions $V^{ml}_{nk}<0$ with $V^{ml}_{nk}$ vanishing when $m=l$ and $n=k$ as guaranteed by Fermi statistics when $N=1$.

To solve this model we take $N\gg 1$ with $V$ finite. We also work in the imaginary time formalism and work in $B=1$ units. We then employ large $N$ vector model techniques, which have been used to study $O(N)$ vector models \cite{Aharony:2020omh}, Chern-Simons-matter theories \cite{Giombi:2011kc,Halder_2019}, the SYK model \cite{Rosenhaus:2018dtp}, and dipole-symmetric models \cite{jensen2022large}. This is done by performing a ``generalized Hubbard-Stratonovich transformation,'' rewriting the problem in terms of bilocal collective fields $G_{mn}(t,t') \equiv \bar{c}^m_\alpha(t) c_n^\alpha(t') $, and the self energy $\Sigma_{mn}(t,t')$ which enforces this constraint. These collective fields are weakly coupled at large $N$, so solving the large $N$ model amounts to solving their classical equations of motion. We ansatz that these fields are invariant under translations in imaginary time and $y$, implying that 
\begin{align}
\begin{split}
\label{E:SDansatz}
	G_{n,m}(\omega_k,\omega_{k'})& = \beta NG_n(\omega_{k'})\delta_{n,m} \delta_{k,-k'}\,,  
	\\
	 \Sigma_{n,m}(\omega_k,\omega_{k'}) &= \beta \Sigma_n(\omega'_{k'}) \delta_{n,m}\delta_{k,-k'}\,,
\end{split}
\end{align}
where $\omega_k = \frac{\pi}{\beta}(2k+1)$ for $k\in\mathbb{Z}$ is the $k^{\rm th}$ Matsubara mode. The large $N$ Dyson equations then are
\begin{equation}
    \begin{split}
        G_n(\omega_k) &= \frac{1}{i\omega_k + \Sigma_n(\omega_k)}, \\
        \Sigma_n(\omega_k) &= -\mu + \frac{2}{\beta} \sum_{k'\in\mathbb{Z}} \sum_{m\in\mathbb{Z}} V_{n,m} G_m(\omega_{k'})\,,
    \end{split}
\end{equation}
where $V_{m,n} = V_{n m}^{n m}$. Note that $V_{m,n}$ is only a function of $|n-m|$, due to the symmetry under magnetic translations in the $x$ direction. These coefficients $V_{m,n}$ are precisely the part of $V^{mn}_{pq}$ that survive in the thin torus limit of~\cite{Tau_Thouless,Bergholtz_QHTT}. Here they govern the problem at finite radius but large $N$. Now, the second equation implies that $\Sigma_n$ is frequency-independent. The two Dyson equations can then be combined into a single equation for the self energy
\begin{equation} 
\label{eq:dyson_sigma}
    \Sigma_n = -\mu + \sum_{m\in \mathbb{Z}} V_{n,m} \tanh\left(\frac{\beta \Sigma_m}{2} \right)\,.
\end{equation}

The large $N$ free energy evaluated on the solution to the the Dyson equations is 
\begin{equation}
\begin{split}
    \frac{F}{ N  L } = - \frac{1}{L}\sum_{n\in \mathbb{Z}} \bigg[ &\frac{1}{\beta} \ln \left(\cosh\left(\frac{\beta\Sigma_{n}}{2} \right) \right)
    \\ 
    & - \frac{1}{4} (\Sigma_n + \mu )  \tanh\left(\frac{\beta\Sigma_{n}}{2}\right) \bigg] \,,
\end{split}
\end{equation}
where $L$ is the length of the cylinder in units of magnetic translations, or equivalently $L$ is the IR cutoff for the sum over $n$. The maximum possible number of filled states is $NL$. From the free energy we can calculate the filling fraction, $\nu$, corresponding to the solution to be \begin{equation}
    \nu=  \frac{1}{2} - \frac{1}{2 L} \sum_{n\in\mathbb{Z} } \tanh\left(\frac{\beta \Sigma_n}{2}\right)\,.
\end{equation}

\textit{Large $N$ solutions: FQH hierarchies and transitions.}~As this setup has a $\mathbb{Z}$ symmetry of magnetic translations along the cylinder, it is natural to look for solutions to the Dyson equations that preserve this symmetry. Thus we can take a translationally invariant ansatz $\Sigma_n = \Sigma$, a constant value, which when nonzero signals a gap in the single-particle spectrum. Using this ansatz we see that there is a unique translationally invariant solution to the Dyson equations \eqref{eq:dyson_sigma}  given by 
\begin{equation}
    \Sigma = -\mu + V \tanh\left( \frac{\beta \Sigma}{2}\right)\,,
\end{equation}
where $V= \sum_{n\in \mathcal{Z}} V_{m,n}<0$ is an $m$-independent constant, as a result of the interaction matrix being a function of $|n-m|$. At small but nonzero temperature this solution is, for $|\mu|>|V|$, $\Sigma \approx -(\mu  + \text{sgn}(\mu)V)$, while for $|\mu|<|V|$ it is $\Sigma \approx \frac{2}{\beta}\text{arctanh}\left( \frac{\mu}{V}\right)$. The free energy and filling fraction for this solution are simply 
\begin{align}
	\nonumber
	\frac{F}{ N  L } &= - \frac{1}{\beta} \ln \left(\cosh\left(\frac{\beta\Sigma}{2}\right) \right) + \frac{1}{4} (\Sigma + \mu )  \tanh\left(\frac{\beta\Sigma}{2}\right)\,,
	\\
 	\nu&= \frac{1}{2}-\frac{1}{2}\tanh\left(\frac{\beta \Sigma}{2}\right)\,.
\end{align}
As zero temperature this solution has $\nu =0$ for $\mu<-|V|$, $\nu=1$ for $\mu > |V|$, and $\nu$ linearly increasing in the region $\mu \in (-|V|,|V|)$. This describes two gapped phases for $|\mu|>|V|$, completely empty and completely filled, continuously connected to each other by a gapless phase, whose existence is due to the interaction.

We also look for solutions to the Dyson equations which only preserve a $q\mathbb{Z}$ subgroup of the translations, i.e. for which magnetic translations are broken to the subgroup generated by translation by $q$ units. Such a solution obeys $\Sigma_{n+q}= \Sigma_{n}$, and is characterized by $q$ undetermined numbers $\{\Sigma_0,\Sigma_1,\ldots,\Sigma_{q-1}\}$ which comprise a unit cell. On such an ansatz the Dyson equation~\eqref{eq:dyson_sigma} becomes
\begin{equation} \label{eq:dyson_modp}
    \Sigma_n = - \mu + \sum_{m=0}^{q-1} V_{n,m}^{(q)} \tanh\left( \frac{\beta \Sigma_m}{2}\right),
\end{equation}
where $V_{n,m}^{(q)} = \sum_{k\in \mathbb{Z}} V_{n,m+kq}$. The free energy and filling fraction for these solutions are  
\begin{align}
\begin{split}
	\frac{F}{ N  L } & =- \frac{1}{p} \sum_{n=0}^{q-1} \bigg[ \frac{1}{\beta} \ln \left(\cosh\left(\frac{\beta\Sigma_n}{2}\right) \right) 
	\\ 
	&\qquad \qquad \qquad  - \frac{1}{4} (\Sigma_n + \mu )  \tanh\left(\frac{\beta\Sigma_n}{2}\right) \bigg] \,,
	\\
	 \nu&= \frac{1}{2}-\frac{1}{2q}\sum_{n=0}^{q-1} \tanh\left(\frac{\beta \Sigma_n}{2}\right)\,.
\end{split}
\end{align}
At values of $\mu$ where more than one solution exists, we compare the free energy of the various solutions to find which one is dominant.

\begin{figure}
    \centering
    \includegraphics[width = \linewidth]{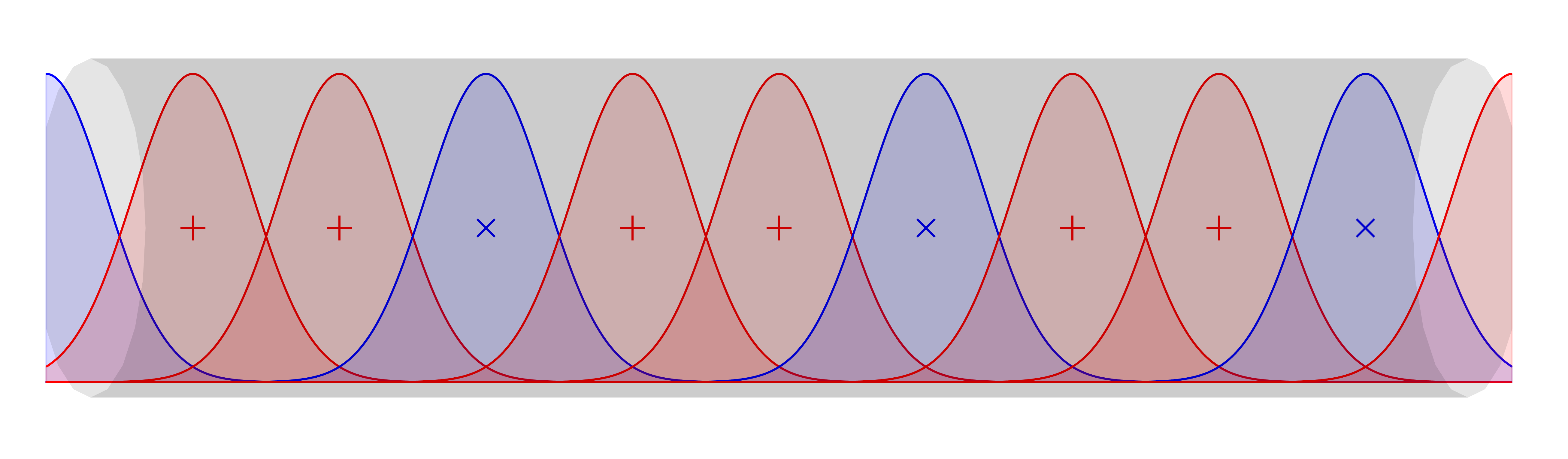}
    \caption{A schematic of a solution to the Dyson equations with $q=3$ and $p = 1$ corresponding to a state of filling fraction $\frac{\nu}{N} = 1/3$. The sites and their corresponding single particle wave-functions on the cylinder are presented, with the blue sites being "filled" and having $\Sigma_n < 0$ and red sites being "empty" and having $\Sigma_n > 0$.}
    \label{fig:cylinder}
\end{figure}

This set of equations, \eqref{eq:dyson_modp}, has a trivial solution where all the $\Sigma_n$ are equal, the solution invariant under magnetic translations that we discussed above. However, and crucially, we also find non-trivial solutions which at zero temperature correspond to gapped, fractionally filled states with $\nu = \frac{p}{q}$ where exactly $p$ of the different $\Sigma_n$'s in a unit cell of length $q$ are negative and the remaining are positive. A schematic picture of these solutions is presented in figure \ref{fig:cylinder}. We have found these solutions numerically for a number of different interactions as we describe shortly. Here let us focus on the general picture. 

These non-trivial solutions only exist for a finite range of chemical potential in the region $\mu \in (-|V|,|V|)$ where they compete with the translationally-invariant gapless phase identified above. However we find that these gapped fractionally filled states have lower free energy than the gapless one. The ensuing phase diagram at zero temperature then interpolates between the translation-invariant fully empty state for $\mu<-|V|$ and the fully filled state for $\mu>|V|$, by way of a series of first order transitions between fractionally filled states.

Fractional filling is not by itself sufficient to demonstrate that these are fractional Hall states. To demonstrate the latter, we first note that these non-trivial solutions in general have degeneracy $q$ as a consequence of the spontaneously broken magnetic translation symmetry. Given a non-trivial solution with periodicity $q$, we obtain $q-1$ more by mapping $\Sigma_n\rightarrow \Sigma_{n+m}$ for $m=1,..,q-1$. Additionally, there are anyonic excitations above these states, domain walls between these different solutions, which themselves can condense and so generate a hierarchy of FQH states as in~\cite{PhysRevLett.51.605,PhysRevLett.52.1583}. 

These anyons, their charges, and their condensation can be understood in a very similar way to that of the thin torus limit as explained in~\cite{Bergholtz_QHTT}. Consider a unit cell $\{\Sigma_0,\hdots,\Sigma_{q-1}\}$ of a non-trivial solution. Define a domain wall to be a set of $m$ self-energies $\mathcal{D} = \{ \widetilde{\Sigma}_0,\hdots ,\widetilde{\Sigma}_{m-1}\}$. Inserting this domain wall into a solution generates a new configuration, not a solution to the Dyson equations~\eqref{eq:dyson_sigma}, that can be interpreted as a quasiparticle excitation. On either side of the domain wall one has one of the $q$ degenerate configurations of the $\Sigma_n$'s that on their own correspond to a genuine solution to the Dyson equations. The charge of these domain walls can be calculated using the counting trick of~\cite{Schrieffer81}. Insert $q$ widely separated domain walls into the chain, with an integer number of unit cells between each, and remove $m$ unit cells from the end so as to maintain $L\gg 1$ sites. The ensuing configuration can be interpreted as a single solution, punctured by an interpolating region where the domain walls are inserted. Accounting for the fact that a single negative $\Sigma_n$ contributes $N$ to the charge and a positive one $0$, the total change in the charge relative to the original configuration encodes a fractional domain wall charge $Q_{\mathcal{D}} = N(n_{\mathcal{D}} - m \nu)$ where $n_\mathcal{D}$ is the number of negative $\widetilde{\Sigma}$'s in $\mathcal{D}$. When $p$ and $q$ are coprime the elementary fractional charge is $\frac{N}{q}$. We can also condense these domain walls to get new states by enlarging the unit cell, defining a new one formed from some number of original unit cells and one domain wall, together with a suitable adjustment of the $\Sigma$'s in the cell (without changing their sign) so that it is a solution to the Dyson equations for some chemical potential. This is similar to the procedure used in \cite{Bergholtz_QHTT} to generate the Haldane-Halperin hierarchy of abelian FQH states.

\begin{figure*}
    \centering
    \includegraphics[width = 0.9\linewidth]{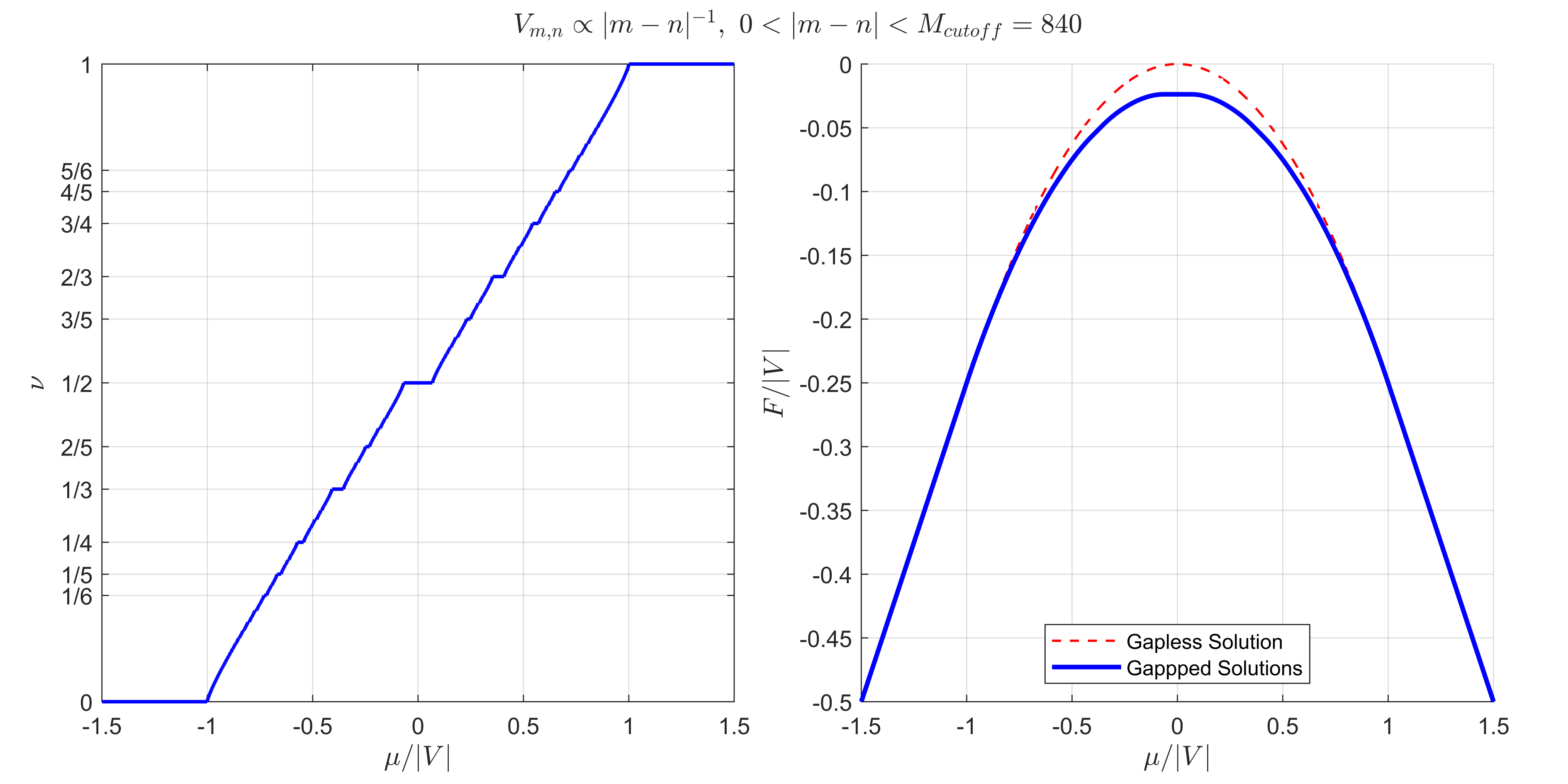}
    \caption{A plot of the filling fraction and free energy as a function of chemical potential at zero temperature for $N$ species of electrons in the lowest Landau level with an interaction matrix $V_{m,n} \propto -|n-m|^{-1}$.}
    \label{fig:coulomb}
\end{figure*}

For an example consider a Coulomb like-interaction between the sites, and take $V_{m,n} \propto -|m-n|^{-1}$ together with $V_{m,m}=0$ and a cutoff on the maximum possible $|m-n|$. We solve the Dyson equations numerically to find the filling fraction as a function of chemical potential. A plot of these solutions at zero temperature as well as the free energy is presented in figure \ref{fig:coulomb}. In this figure we can see the series of transitions between the translationally invariant states for $|\mu| >|V|$ through a sequence of fractionally filled states, where states with a larger denominator in the filling have larger plateaux. For this long range interaction the plateaux transitions seem to form a ``devil's staircase'' with a self-similar structure as the cuttoff is increased, however this behavior isn't observed for short range interactions where solutions with a periodicity that is much larger than the interaction length do not exist.\footnote{This is evident for (admittedly unrealistic) interactions $V_{n,m}$ with a hard cutoff whereby $V_{n,m}=0$ for $|n-m|$ above the cutoff. For such an interaction there are solutions with a periodicity $q$ only for $q$ less than the cutoff. That is, a hard cutoff in the interaction imposes a hard cutoff on $q$.} In general we do get both plateaux where $p$ and $q$ can be even or odd, although this may be a large $N$ artifact as we discuss later. We elaborate more on how we obtains these numerical solutions in the Appendix, and showcase the phase diagram for a short-range interaction as well.

\textit{Extrema at $N=1$.}~While we arrived at the Dyson equations from standard large $N$ arguments, some  of our analysis holds at finite $N$, and even at $N = 1$. We can exactly rewrite the original problem in terms of bilocal degrees of freedom $G$ and $\Sigma$ without introducing any Jacobian, so the large $N$ action and its equations of motion for $G_{n,m}$ and $\Sigma_{n,m}$ hold even at finite $N$.

On our ansatz~\eqref{E:SDansatz} the parameter $N$ completely drops out of those equations of motion, and therefore the solutions we identified above at large $N$ are in fact extrema of the many-body action at any $N$ including $N=1$. However at $N=1$ the second-quantized path integral no longer has a saddle-point approximation. Fluctuations of the bilocal degrees of freedom around the extrema have $O(1)$ interactions at finite $N$. Even so, to the extent that it makes sense to organize the interactions into a perturbative expansion, the filling fractions corresponding to these extrema are exact to all orders in perturbation theory by the following argument. Loop effects will generate frequency-dependent radiative corrections to the self-energies $\Sigma_n$, but these effects are perturbatively small and therefore do not change the sign of the $\Sigma_n$. However, at zero temperature, the filling fraction is solely due to the signs of the $\Sigma_n$.

While these solutions persist at finite $N$ and their filling fraction is exact in perturbation theory, their free energy receives $O(1)$ corrections at finite $N$. This means the precise phase structure and transitions between the plateaux at finite $N$ will differ from our large $N$ results, as different states become energetically favored as a function of $N$ and $\mu$. In general we expect the gapped even denominator states to become less energetically favorable, as the observed states at even denominator filling are gapless. This is similar to the TT picture of \cite{Bergholtz_QHTT}, where the gapped even denominator states become energetically suppressed away from the thin torus limit.

\textit{Discussion.}~In this Letter we have provided a first-principles derivation of fractional Hall states and transitions from a microscopic description of strongly interacting fermions in the lowest Landau level using large $N$ techniques. One big advantage of our approach is the ability to study many fractional quantum Hall states simultaneously, from a single Hamiltonian, and to observe Hall plateaux without any need for disorder. This framework can also be used to study other aspects of the fractional quantum Hall system, including finite temperature effects, the nature of plateau transitions, and the conductivity of each state. Another advantage of a large $N$ approach is the ability to systematically compute the corrections to the leading large $N$ results.

Unfortunately our approach has one big drawback: the existence of even denominator plateaus. These even denominator gapped states may be large $N$ artifacts, and we expect them to be subdominant to a gapless phase at $N=1$.  More generally, our solutions exist as extrema of the many-body problem even at finite $N$, though they receive energetic corrections of $O(1/N)$, thereby invalidating our phase diagram. Nevertheless their filling/degeneracy persist even at $N=1$.

Our methods can in principle be adapted to study the problem of $N\gg 1$ species of interacting fermions in the LLL on other symmetric spatial geometries, namely the sphere, the plane, and the torus. We are doing so presently and hope to report our findings soon.

To conclude we note that there is a controlled way to use the large $N$ theory to obtain abelian FQH states that resemble those expected for the physical case of a single species of interacting fermion. We simply project the large $N$ theory onto its $SU(N)$ singlet sector. In the second quantized description this can be accomplished by coupling the matter to a $SU(N)_k$ Chern-Simons theory in the $k\to \infty$ limit. Crucially our solution of the unprojected theory is also the solution of the projected one, which only has a $U(1)$ global symmetry. After this projection the elementary fermions $c_n^{\alpha}$ are unphysical, and the basic operators that carry $U(1)$ charge are ``baryons,'' like $\epsilon_{\alpha_1\hdots \alpha_N} c_n^{\alpha_1}\hdots c_n^{\alpha_N}$, whose $U(1)$ charge is quantized in multiples of $N$. Rescaling the $U(1)$ charge so that baryons have charge $1$, i.e. charge is quantized in multiples of 1, the states we find have all of the expected data for FQH states built from the LLL. Namely, a filling fraction $\frac{p}{q}$ with degeneracy $q$ on the torus, and fractionally charged quasiparticles of charge $\frac{\text{gcd}(p,q)}{q}$.

\emph{ Acknowledgments.}
We would like to thank A.~Abanov, A.~Karch, T.~Grover, A.~Gromov, M.~Ippoliti, A.~MacDonald, M.~Mezei, D.~T.~Son, and D.~Tong for enlightening discussions. KJ is supported in part by an NSERC Discovery Grant, and AR is supported by the U.S. Department of Energy under Grant No. DE-SC0022021 and a grant from the Simons Foundation (Grant 651678, AK).

\bibliographystyle{apsrev4-1}
\bibliography{refs}

\onecolumngrid

\appendix

\section{Details of the numerical solutions} \label{app:numrics}

In this appendix we briefly describe our algorithm for numerically solving the Dyson equations~\eqref{eq:dyson_modp}. For simplicity all of the solutions we found were gapped at zero temperature, i.e. they are described by a set of nonzero self-energies $\Sigma_n$. We worked in units where $V = \sum_{n\in \mathbb{Z}} V_{m,n} = -1$, so that the range of chemical potentials $\mu$ where we find non-trivial phases is in the interval $\mu\in [-1,1]$.
 
 At zero temperature the $\tanh()$ becomes a $\text{sgn}()$ function, so it much easier to fix a vector $s_n = \text{sgn}(\Sigma_n)$ and then find the range of $\mu$'s where the Dyson equations have a consistent solution. In particular this range is $\mu \in (\mu_{\min},\mu_{\max})$, where
\begin{equation}
\begin{split}
    \mu_{\min} &= \max_{\{m \in \mathbb{Z}_q : s_m = -1\}} {\left(\sum_{n = 0}^{q-1} V_{m,n} s_n\right)}\,,
    \\
    \mu_{\max} &= \min_{\{m \in \mathbb{Z}_q : s_m = +1\}} {\left(\sum_{n = 0}^{q-1} V_{m,n} s_n\right)}\,.
\end{split}
\end{equation}
Then we compute the free energy~\eqref{eq:dyson_modp} for this solution in the allowed range and compare it to other allowed solutions to find the dominant phase. We also compare it to the transitionally invariant gapless phase where $\Sigma = 0$ but $\tanh(\beta \Sigma/2) \rightarrow s \in [-1,1]$.

\begin{figure*}
    \centering
    \includegraphics[width = 0.9\linewidth]{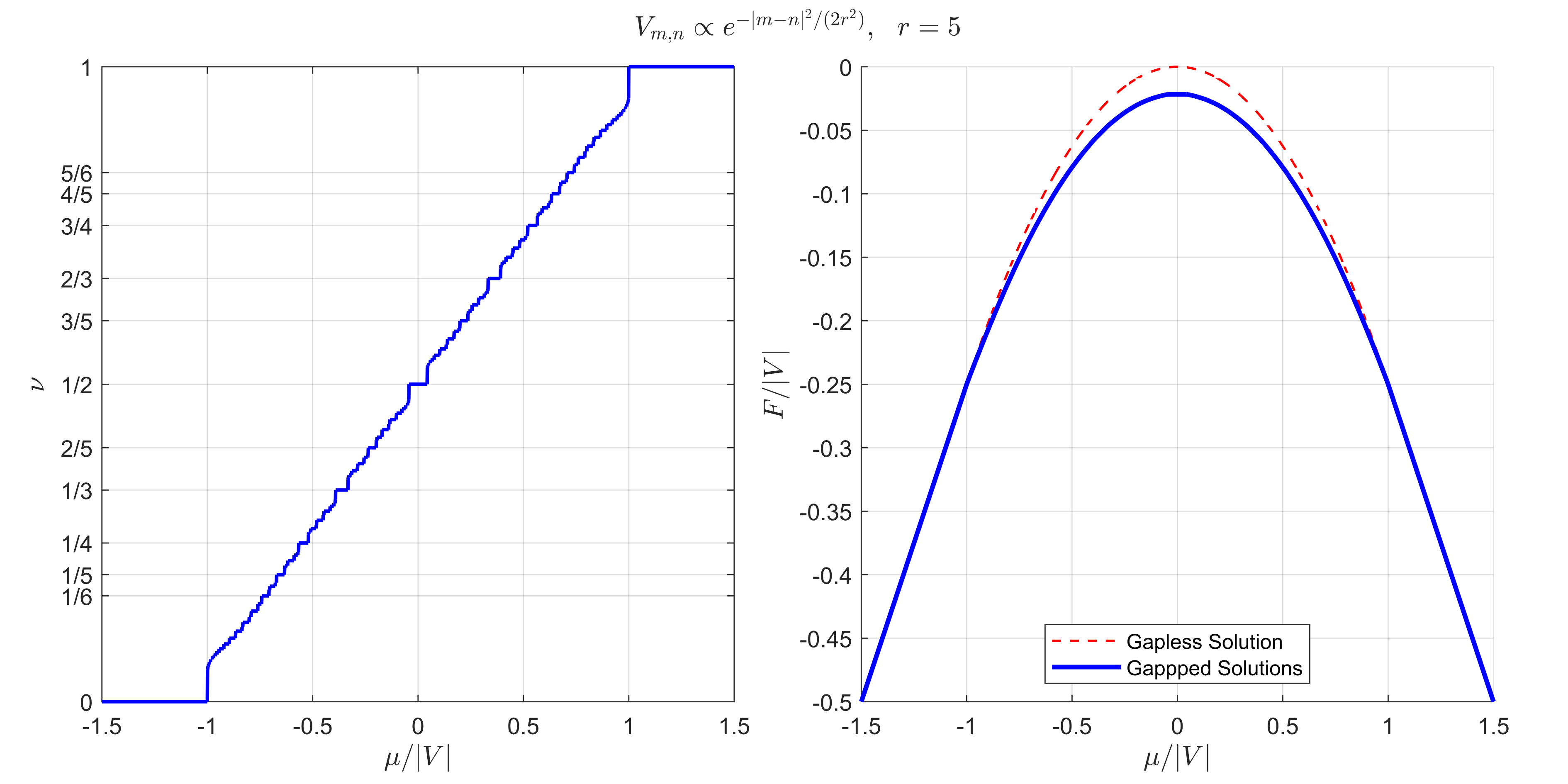}
    \caption{A plot of the filling fraction and free energy as a function of chemical potential at zero temperature for electrons in the lowest Landau level with an interaction matrix $V_{m,n} \propto e^{-|n-m|^2/(2r^2)}$ together with $V_{m,m}=0$, with radius $r=5$. This interaction matrix arises from a local four-fermion interaction term.}
    \label{fig:local}
\end{figure*}

Unfortunately the number of possible sign strings $\{s_n\}$ grows exponentially with $q$, so it is only possible to check all possible strings for $q\lesssim 17$. We can check solutions with a much higher values of $q$ if we only restrict ourselves to certain strings $\{s_n\}$. One natural choice of strings are the most diffusive strings, which are also the TT states considered in \cite{Bergholtz_QHTT}. This configuration is the unique string of length $q$ (up to translations mod $q$,) with $p$ $s_n$'s being $-1$, and where these $-1$ values are the most  spread out. The number of these strings for a fixed $q$ is proportional to $q$, so considering all such strings up to some maximal length $q_{\max}$ requires only $O(q_{\max}^2)$ solutions to check. Checking all such strings can be done up to $q_{\max} \sim  400$ in a reasonable amount of time.This is how figure \ref{fig:coulomb} was created, as well as figure \ref{fig:local}, which shows a similar plot for a local interaction term.

\begin{figure*}
    \centering
    \includegraphics[width = 0.9\linewidth] {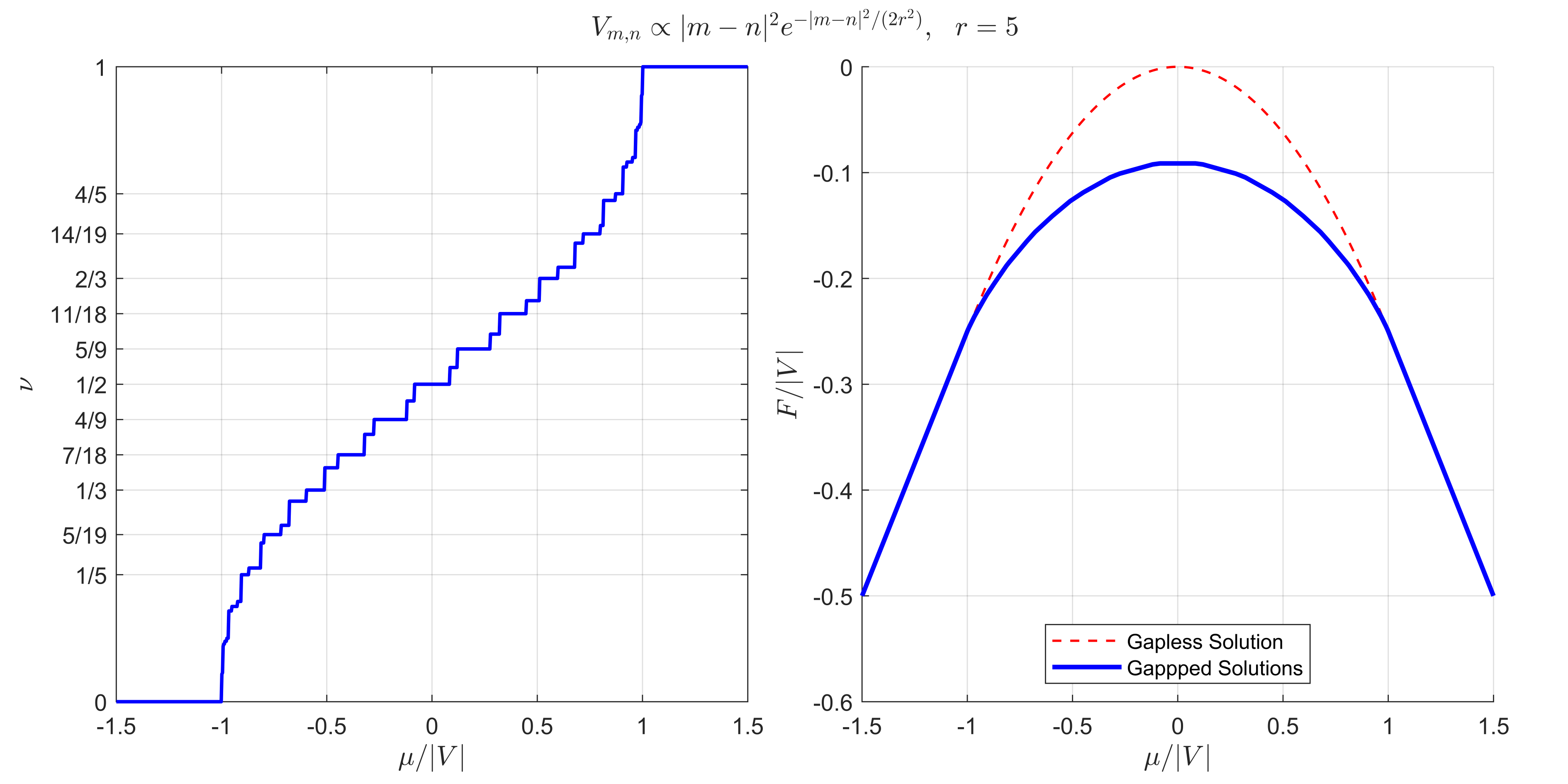}
    \caption{\label{fig:localderivint} A plot of the filling fraction and free energy as a function of chemical potential at zero temperature for electrons in the lowest Landau level with an interaction matrix $V_{m,n} \propto |n-m|^2 e^{-|n-m|^2/(2r^2)}$ and radius $r = 5$.}
\end{figure*}

Unfortunately not all interesting interaction terms lead to these simple TT-like solutions. For example the interaction term from \cite{nguyen2014lowest}, $S_{\text{int}} = - \lambda \int |\psi D \psi|^2$, leads to the interaction matrix $V_{m,n} \propto -|n-m|^2 e^{-|n-m|^2/(2r^2)}$. For this interaction there is a small energy cost for neighboring sites to both be occupied, and a much larger energy cost for sites $\sim \sqrt{2} r$ apart of being occupied. Thus, for this interaction the TT-like diffusive states aren't dominant, and a different class of strings should be checked. For this type of interaction we looked at strings with a limited number of sign changes between $s_n$ and $s_{n+1}$. We checked all strings of length 50 with fewer than five sign changes, strings of length 30 with fewer than 8 sign changes, stings of length 25 with fewer than 10 sign changes. A plot of the filling fraction and free energy for this interaction is presented in Fig.~\ref{fig:localderivint}.

\end{document}